# Dispersion-Engineered Compact Twisted Metasurfaces Enabling 3D Frequency-Reconfigurable Holography

Cheng Pang[1†], Yuzhong Wang[1†*], Pengcheng Wang[1], Axiang Yu[1], Yiding Liu[1], Ziang Yue[1], Mingshuang Hu[1], Jianqi Hu[3], Yongkang Dong[2] and Jiaran Qi[1*]

[1] Department of Microwave Engineering, School of Electronics and Information Engineering, Harbin Institute of Technology, Harbin, 150001, China
[2] National Key Laboratory of Laser Spatial Information, Harbin Institute of Technology, Harbin, 150001, China
[3] Institute of Physics, Swiss Federal Institute of Technology Lausanne (EPFL), Lausanne, 1015, Switzerland

*Correspondence: hitwyz@stu.hit.edu.cn; qi.jiaran@hit.edu.cn;
†These authors contributed equally to this work.

**Abstract.** Flexible dispersion manipulation is critical for holography to achieve broadband imaging or frequency division multiplexing. Within this context, metasurface-based holography offers advanced dispersion control, yet dynamic reconfigurability remains largely unexplored. This work develops a dispersion-engineered inverse design framework that enables 3D frequency-reconfigurable holography through a twisted metasurface system. The physical implementation is based on a compact layered configuration that cascades the broadband radiation-type metasurface (RA-M) and phase-only metasurface (P-M). The RA-M provides a phase-adjustable input to excite P-M, while the rotation of P-M creates a reconfigurable response of holograms. By employing the proposed scheme, dynamic switching of frequency-space multiplexing and achromatic holograms are designed and experimentally demonstrated in the microwave region. This method advances flexible dispersion engineering for metasurface-based holography, and the compact system holds significant potential for applications in ultra-broadband imaging, high-capacity optical display, and switchable meta-devices.

## Introduction

Holography, which utilizes optical interference and diffraction to reconstruct three-dimensional (3D) objects, has gained significant attention in advanced photonics research [1-5]. One of the desirable features of holographic techniques is the ability to flexibly manipulate the dispersion of reconstructed fields [6-10]. For instance, near-eye displays [11-12] demand chromatic aberration elimination to improve imaging quality, while color holographic displays [13-14] require flexible control of dispersion characteristics for dynamic spectral tuning. Traditional approaches rely on cascaded optical elements (e.g., lenses and stacking polarizers) to realize the holographic dispersion control. However, these methods typically escalate structural complexity and cannot be miniaturized within compact optoelectronic platforms.

Recently, metasurface-based holography has emerged as a revolutionary alternative [15]. It leverages ultra-compact optical elements on a two-dimensional plane to manipulate the properties of electromagnetic (EM) in a point-by-point manner [16-20]. In particular, the advancements in metasurface platforms have demonstrated unprecedented control over dispersion characteristics. Among them, most efforts have been made to eliminate the negative chromatic aberration in typical meta-atoms21, using sub-elements [22], interleaved element arrangement [23], and multi-layer elements [24]. These techniques mainly focus on structural design and theoretical approaches to construct dispersion-engineered meta-atoms [25]. By constructing the meta-atoms library with rich dispersion responses, the desirable holographic dispersion control has been achieved across various frequencies [26-27]. Further, the linear phase compensation method is applied to approximate the dispersion profiles [28-32], which can eliminate chromatic aberration in a certain operating band and thus improve the imaging quality. Meanwhile, the discrete wavelength dispersion control is realized via the structural dispersion design freedom of subwavelength elements, enabling wavelength-multiplexed holography [33-34]. Such multi-dimensional multiplexed designs facilitate diverse emerging applications35-36, such as full-color holography that increases information display and storage capacity [37-38]. Overall, the metasurface platform provides remarkable flexibility in holographic dispersion engineering.

Despite tremendous progress in past years, metasurface-based holographic designs generally have fixed dispersion characteristics. While this is beneficial for specific single-application scenarios such as security surveillance [39], reconfigurable dispersion control is essential for versatile applications such as optical encryption [40] and dynamic displays [41], where greater flexibility can enhance the channel capacity. Current reconfiguration strategies primarily focus on input wavefront modulation [42-43] or tunable meta-atoms [44-45]. The former approach uses spatial light modulators (SLMs) or other optical elements to achieve dynamic holographic imaging through intensity [46-47] or polarization [48-49] modulation. The latter exploits dynamic materials to construct the meta-atoms and achieve switching functions via external control devices. However, existing meta-devices often overlook the reconfigurability of their dispersion. This is since the dispersion is linked to the group delay characteristics of meta-atoms, which are usually determined by their structural design. Common building blocks like active components, liquid crystals, and phase change materials, primarily control phase [50-51], amplitude [52], and polarization [53-54] rather than dispersion properties. As such, simple and effective reconfigurable dispersion control of meta-atoms remains a challenge.

In this work, we propose a dispersion-engineered inverse design framework (DIDF) to achieve frequency-reconfigurable metasurface holograms. DIDF is based on the layered configuration of cascaded broadband radiation-type metasurface (RA-M) and phase-only metasurface (P-M), forming a compact twisted metasurface system. The reconfigurable response of the metasurfaces is obtained by the rotation of the P-M while keeping the RA-M fixed. The RA-M creates a modulated wavefront for the P-M to excite the corresponding EM response coded in the rotational P-M, realizing dynamic switching of frequency-reconfigurable holograms. By constructing the relationship between the target dispersion-specific holograms and the phase profile of RA-M and P-M, high-quality image reconstruction of the frequency-reconfigurable holograms in the 3D space is achieved. For validation, we experimentally demonstrate the dynamic switching of space-frequency multiplexing and achromatic holograms based on the proposed framework in the microwave region. This framework advances flexible dispersion engineering in metasurface-based holography and motivates further development of dispersion-engineered meta-devices for other practical applications.

**Methods**

Here, we use DIDF to map the frequency-specific holographic information into the EM response of the twisted metasurface system. The operating mechanism of DIDF in a layered configuration of metasurfaces is illustrated in Fig. 1. Two independent metasurfaces, i.e., RA-M and P-M are assembled into a compact meta-device. The broadband RA-M is excited by an integrated feeding network while the P-M is excited by the RA-M, thus the whole system is free from any external bulky feeds. By tuning the rotatable P-M at a specific twisted angle, the broadband EM signal generated by the RA-M is modulated with specific phase distribution and interacts with P-M. Complex diffraction occurs inside the layered configuration to generate a unique EM response and finally reconstruct the holographic images at specific output planes. Fig. 1 also showcases an illustration of the process. Three types of dispersion-engineered holograms are illustrated, covering frequency multiplexing, space-frequency multiplexing, and chromatic aberration correction, all mapped into the twisted operation (e.g., characters "1" "2" and "3" at 13 GHz, 15 GHz, and 17 GHz of z1 plane for state 1; characters "4" at 13 GHz of $z_1$ plane, characters "5" at 15 GHz of $z_2$ plane, characters "6" at 17 GHz of $z_3$ plane for state 2; characters "7" at 12 – 18 GHz of $z_1$ plane for state 3. Through the dynamic switching of the P-M, complex holographic images are then updated. Notably, the twisted operation is extensively adaptable and can be used in diverse application scenarios.

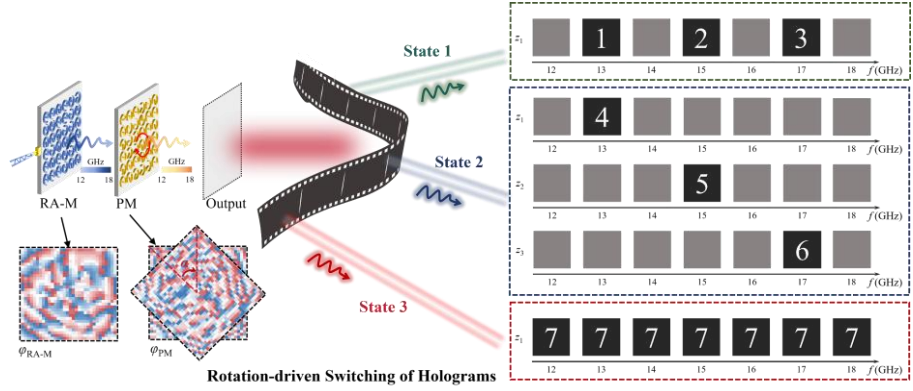

**Fig. 1** Architecture of the proposed compact twisted meta-device for frequency-reconfigurable holography. The meta-device integrates cascaded radiation-type metasurfaces (RA-M) with phase-only metasurfaces (P-M), where in-plane rotational control of the P-M through angle α enables dynamic hologram switching across designated imaging planes.

To further understand the working principle and realization of DIDF, we start with a fundamental diffraction process of the twisted metasurface system. The EM signal is first modulated by the RA-M and creates a wavefront $U_{RA-M}(x, y, f)$. Then, the $U_{RA-M}$ ($x, y, f$) propagates a certain distance inside the layered configuration to $\mathbf{U}_{\text{Inc}}$ ($x, y, f$) and then interacts with P-M. The P-M featured with $\mathbf{U}_{\text{P-M}}$ ($x, y, f$) further modulates the EM signal, ultimately rendering intricate diffractive effects at the output plane. The reconstructed field $\mathbf{U}_{\text{Image}}$ ($x, y, z, f$) at the output plane can be expressed as:

$$\mathbf{U}_{\text{Inc}}(x, y, z, f) = \iint \mathbf{U}_{\text{RA-M}}(x_0, y_0, f) h(x - x_0, y - y_0, z, f) dx_0 dy_0$$

$$\mathbf{U}_{\text{Image}}(x, y, z, f) = \iint \mathbf{U}_{\text{Inc}}(x, y, z, f) \mathbf{U}_{\text{P-M}}(x_0, y_0, f) h(x - x_0, y - y_0, z, f) dx_0 dy_0$$

(1)

Where $h$ ($x, y, z, f$) is the impulse response. Based on Equation (1), $\mathbf{U}_{\text{Image}}$ ($x, y, z, f$) can be controlled by the distribution of $\mathbf{U}_{\text{RA-M}}$ ($x, y, f$) and $\mathbf{U}_{\text{P-M}}$ ($x, y, f$), and the joint operation of $\mathbf{U}_{\text{RA-M}}$ ($x, y, f$) and $\mathbf{U}_{\text{PM}}$ ($x, y, f$) will yield distinct EM functionalities on the output plane. Considering phase-only modulation, $\varphi_{\text{RA-M}}$ ($x, y, f$) and $\varphi_{\text{PM}}$ ($x, y, f$) can finally contribute to the $\mathbf{U}_{\text{Image}}$ ($x, y, z, f$), and both two components have similar expressions:

$$\varphi_{\text{RA-M}}(f) = \begin{bmatrix} \varphi_{11}^{\text{RA-M}}(f) & \cdots & \varphi_{1N}^{\text{RA-M}}(f) \\ \cdots & \varphi_{ij}^{\text{RA-M}}(f) & \cdots \\ \varphi_{N1}^{\text{RA-M}}(f) & \cdots & \varphi_{NN}^{\text{RA-M}}(f) \end{bmatrix}^{N \times N}$$

$$\varphi_{\text{P-M}}(f) = \begin{bmatrix} \varphi_{11}^{\text{P-M}}(f) & \cdots & \varphi_{1N}^{\text{P-M}}(f) \\ \cdots & \varphi_{ij}^{\text{P-M}}(f) & \cdots \\ \varphi_{N1}^{\text{P-M}}(f) & \cdots & \varphi_{NN}^{\text{P-M}}(f) \end{bmatrix}^{N \times N}$$

(2)

where $N$ is the number of the discrete points along the $x$- and $y$- directions. Based on the Rayleigh-Sommerfeld diffraction theory , each discrete point can be considered as an independent secondary source, thus an extra twisted angle $\alpha$ can be added to Equation (2) to change the effective interaction region between $\varphi_{\text{RA-M}}$ ($x, y, f$) and $\varphi_{\text{P-M}}$ ($x, y, f$). Here, we apply the twisted operation to $\varphi_{\text{P-M}}$ ($x, y, f$), and as the twisted angle $α$ changes, the output $\mathbf{U}_{\text{Image}}$ $(x, y, z, f)$ renders a set of reconstructed images to realize dynamic switching of frequency-reconfigurable holograms. On this basis, the DIDF is applied to optimize the phase profile of both $φ_{\text{RA-M}}$ $(x, y, f)$ and $φ_{\text{P-M}}$ $(x, y, f)$ to obtain the desired holographic images at the target image plane and operating frequency, and finally, establishing the frequency-on-demand evolution process to realize the high-purity reconstruction of the dispersion-specific image switching. The detailed flowchart of DIDF is depicted in Fig. 2. Firstly, the input of DIDF is the phase distribution of $φ_{\text{RA-M}}$ and $φ_{\text{P-M}}$ that are randomly generated and the $\mathbf{U}_{\text{RA-M}}$ $(x, y, f)$ is then obtained. The radiation wavefront of the RA-M experiences the wave free-space propagation via the RSDT to serve as the input of the P-M, which reads

$$w_m\left(x, y, z, f\right) = \frac{z - z_m}{r^2}\left(\frac{1}{2\pi r} + \frac{1}{j\lambda}\right)\exp\left(\frac{j2\pi r}{\lambda}\right) \qquad (3)$$

where $w_m$ is the complex field to connect the front layer with the $m$-th elements located at ($x_m$, $y_m$, $z_m$) of the metasurface aperture, $\lambda$ is the operating wavelength, and $r = \sqrt{(x-x_m)^2 + (y-y_m)^2 + (z-z_m)^2}$. According to the preset twisted angle states matrix $\mathbf{R} = [\alpha_1 \quad \alpha_2 \quad \alpha_3]$, the interaction of the EM waves and the P-M featured with $\mathbf{U}_{\text{P-M}}(x, y, f)$ generates output fields to obtain the intensity distribution at a specific $z$-axis and operating frequency. The disparity between each set of diffraction fields and the targets is characterized by the mean squared error (MSE). Assuming that the output field is discretized as $M \times M$ matrix. For a certain state that operates at $f_q$ GHz and is observed at $z = z_p$ plane, the loss function is defined as the MSE between the output intensity distribution and the target intensity distribution, which can be expressed as:

$$l = \frac{\sum_{m}^{M^2} \left| k I_{\text{Output}}(x_m, y_m, f_q, z_p) - I_{\text{Target}}(x_m, y_m, f_q, z_p) \right|^2}{M^2}$$

$$k = \frac{\sum_{m}^{M^2} I_{\text{Target}}(x_m, y_m, f_q, z_p)}{\sum_{m}^{M^2} I_{\text{Output}}(x_m, y_m, f_q, z_p)}$$

(4)

where $M^2$ is the number of discretized points in the output plane, $I_{\text{Output}}(x_m, y_m, f_q, z_p)$ is the output intensity, $I_{\text{Target}}(x_m, y_m, f_q, z_p)$ is the target normalized intensity, $k$ is the balancing factor used to normalize the target and output field intensities. The gradient descent algorithm and error back-propagation algorithm are used to update the parameters and realize the desired holographic function. Through the iterative process of forward diffraction propagation and error back-propagation, the complete frequency-specific holographic information is encoded into the twisted metasurface system as the progressive reduction in MSE and simultaneous enhancement of output field correlation. We use the Adaptive Moment Estimation Optimizer to optimize the model.

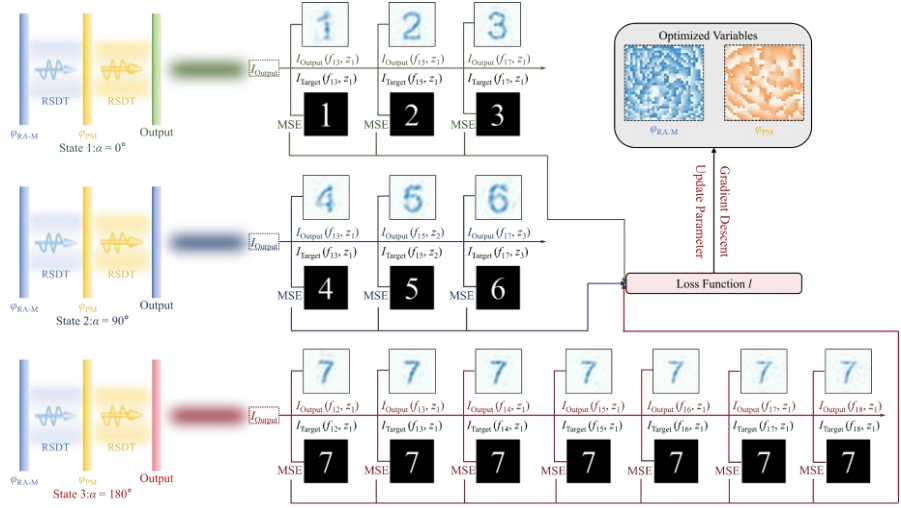

**Fig. 2** The flowchart of the proposed inverse design framework. The input information covers the intensity, frequency, and position of the target imaging plane under different states. The Rayleigh-Sommerfeld diffraction theory is applied to calculate the electric field distributions at each stage. The loss function is used as error evaluation to converge the electric field distribution of the output to one of the presets.

The whole optimization process is implemented in Python v3.8.0 and Pytorch v1.8.0. A personal computer configured with Intel(R) Core (TM) i7-10870H CPU, NVIDIA GeForce RTX 2070 with Max-Q design (GPU), and 16G RAM is used to optimize the models. After the optimization process is finished, the required phase profile $\varphi_{\text{RA-M}}$ and $\varphi_{\text{P-M}}$ of RA-M and P-M is then determined. On this basis, the meta-atoms can be selected from the meta-atom library and further construct the desired dual-layer meta-device.

**Results and Discussion**

To experimentally realize the proposed layered twisted metasurface system, two kinds of meta-atoms are applied to construct the corresponding metasurfaces. The meta-atom for RA-M is shown in Fig. 3(a). This meta-atom is composed of three copper layers and two F4BM-2 dielectric substrates with thicknesses of 1.43 mm and 0.43 mm. The top copper layer patch serves as the EM wave radiator which yields the linearly polarized (LP) wave, while the bottom copper layer is the feeding structure to provide the integrated excitation of the meta-atom, as shown in Fig. 3(b). The middle copper is the shared ground of both the radiation patch and the feeding structure and connects them by the metal holes with a diameter of 0.5 mm. The periodicity of the meta-atom is 6 mm. After optimization, the proposed meta-atom can realize the broadband radiation amplitude above 0.75 from 12 GHz to 18 GHz, as shown in Fig. 3(c). Meanwhile, the length $l_1$ of the bottom-layer feeding lines induces the $2\pi$ propagation phase coverage of the radiation phase, as shown in Fig. 3(d). The optimized parameters of the meta-atom for RA-M are as follows: $r_1$ = 2.1 mm, $\theta_1$ = 40°, $w_1$ = 0.8 mm, $w_2$ = 1.3 mm, $w_3$ = 0.8 mm, $r_2$ = 0.3 mm, $w_4$ = 0.25 mm, $l_2$ = 1.2 mm.

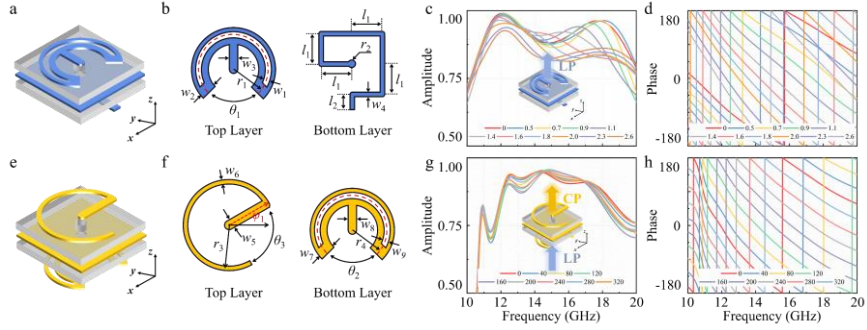

**Fig. 3** a) 3D exploded view of the meta-atom for RA-M. b) Geometric structure of the meta-atom for RA-M. c) Simulated radiation amplitude response of the meta-atom for RA-M versus length $l_1$ of the feeding line. d) Simulated radiation phase response of the meta-atom for RA-M versus length $l_1$ of the feeding line. e) 3D exploded view of the meta-atom for P-M. f) Geometric structure of the meta-atom for P-M. g) Simulated transmission amplitude response of the meta-atom for P-M versus rotation angle $\varphi_1$. h) Simulated transmission phase response of the meta-atom for P-M versus rotation angle $\varphi_1$.

Figure 3(e) depicts the meta-atom of the PM. It consists of three copper layers and two F4BM-2 dielectric substrates with thicknesses of 1.43 mm. The top and bottom copper layers have LP-sensitive and circularly polarized (CP)-sensitive patches, which are used to receive the LP-matched incidence wave from the bottom layer patch and transfer the energy via the metal holes to yield the CP-matched wave from the top layer patch. The detailed structure of the top and bottom patches is shown in Fig. 3(f). The periodicity of the meta-atom is 6 mm. After optimization, the proposed meta-atom can realize the broadband transmission amplitude above 0.8 from 12 GHz to 18 GHz, as shown in Fig. 3(g). Meanwhile, the in-plane rotation $\varphi_1$ of the top-layer patch induces the $2\pi$ phase coverage of the transmission phase, as shown in Figure. 3h. The optimized parameters of the meta-atom for P-M are as follows: $r_3 = 2.3$ mm, $w_5 = 1.3$ mm, $w_6 = 0.3$ mm, $\theta_3 = 20°$, $r_4 = 2.3$ mm, $w_7 = 0.6$ mm, $w_8 = 1.3$ mm, $w_9 = 0.3$ mm, $\theta_2 = 50°$.

To verify the proposed inverse design principle, we conduct the DIDF to physically realize a layered twisted metasurface system for frequency-specific holography. The configuration is composed of 64 × 64 meta-atoms for both RA-M and P-M and occupies the size of 396 mm × 396 mm. We further design a 1-to-4096 feeding network to ensure that the guided EM waves are fed uniformly to each meta-atom of RA-M. Detailed information on the feeding network is shown in Section S1(Supporting Information). The whole thickness of RA-M is 1.965 mm, which corresponds to $0.098\lambda$ at 15 GHz and the one of P-M is 2.965 mm, which corresponds to $0.148\lambda$ at 15 GHz. The RA-M and P-M are placed 100 mm away from each other and the whole dual-layer cascading meta-device occupies a profile along the z-axis of 105.57mm, which corresponds to $5.25\lambda$ at 15 GHz.

A sample with three switching states of P-M for dispersion-customized holography is designed. For state 1 with $\alpha_1 = 0°$, the EM response of the proposed sample shows the multi-frequency holograms in a single plane along the $z$-axis. Three characters "1", "2", and "3" in 13 GHz, 15 GHz, and 17 GHz are selected as the target images located at $z_1 = 150$ mm. When the in-plane rotation of PM turns to $\alpha_2 = 90°$, the EM response of state 2 shows the 3-D multi-frequency holograms. Three characters "4", "5", and "6" in 13 GHz, 15 GHz, and 17 GHz are selected as the target images located at $z_1 = 150$ mm, $z_2 = 200$ mm, and $z_3 = 250$ mm. When the in-plane rotation of PM turns to $\alpha_3 = 180°$, the EM response of state 3 shows the achromatic holograms in a single plane along the $z$-axis. The character "7" in 13 GHz - 18 GHz is selected as the target images located at $z_1 = 150$ mm. We apply the above setting as the target function and develop the proposed DIDF to optimize the phase profile of the RA-M and P-M. The optimized phase distributions of three states are given in Fig. 4(a). The designed meta-device is simulated by the time-domain solver in CST Microwave Studio. The simulated results are shown in Fig. 4(b)-(d). For each switching state, we give the simulated holograms at each frequency ranging from 12 GHz to 18 GHz at the interval of 1 GHz and focal $z$-axis plane ranging from $z_1 = 150$ mm, $z_2 = 200$ mm, and $z_3 = 250$ mm. The intensity distributions in the target frequency are marked with red dashed boxes, while the one of the undesigned frequency is marked with gray dashed boxes. It is noted that the simulated intensity distributions at each target frequency and focal $z$-axis plane agree well with the target image settings. Meanwhile, since the proposed DIDF fails to limit the intensity distribution to a narrow bandwidth, there exists crosstalk observed in the nearby frequency points. The imaging efficiency ($\eta$) of each state is further calculated as

$$\eta = \sum \frac{\mathbf{U}_{out}^2}{\mathbf{U}_{input}^2} \tag{5}$$

where $\mathbf{U}_{input}$ represents the radiation fields of RA-M, $\mathbf{U}_{out}$ represents the output field on the specific $z$-axis plane (imaging plane), and $\sum$ represents the summing of energy on the specific $z$-axis plane. The simulated imaging efficiency for state 1 of "1", "2", "3" are 18.4%, 25.9%, and 37.6%, respectively. The simulated imaging efficiency for state 2 of "4", "5", and "6" are 24.1%, 27.8%, and 29.8%, respectively. The simulated imaging efficiency for state 3 of "7" ranging from 12 GHz to 18 GHz is 14.8%, 16.1%, 16%, 15.7%, 16.3%, and 17.6%, respectively. To evaluate the quality of the reconstructed images by the proposed method, the signal-to-noise ratio (SNR) and Pearson Correlation Coefficient (PCC) is introduced, which reads

$$\text{SNR} = 10\log_{10}\left(\frac{\sum I^2}{\sum (I-O)^2}\right) \tag{6}$$

$$\text{PCC}(I,O) = \frac{\sum (I-\bar{I})(O-\bar{O})}{\sqrt{\sum (I-\bar{I})^2 \sum (O-\bar{O})^2}} \tag{7}$$

where $I$ and $O$ are the intensity of the reconstruction image and target image. We calculated the simulated SNR and PCC of each state at frequency from 12 GHz to 17 GHz and focal z-axis plane versus $z_1 = 100$ mm, $z_2 = 150$ mm, and $z_3 = 200$ mm. The simulated SNR for state 1 of "1", "2", "3" are 2.07 dB, 3.21 dB, and 3.54 dB, respectively. The simulated SNR for state 2 of "4", "5", "6" are 2.89 dB, 3.43 dB, and 3.42 dB, respectively. The simulated SNR for state 3 of "7" ranging from 12 GHz to 18 GHz are 2.90 dB, 3.20 dB, 3.22 dB, 3.09 dB, 2.85 dB, and 2.21 dB, respectively. The simulated PCC for state 1 of "1", "2", "3" are 0.8396, 0.8898, and 0.9000, respectively. The simulated PCC for state 2 of "4", "5", and "6" are 0.8564, 0.8752, and 0.8713, respectively. The simulated PCC for state 3 of "7" ranging from 12 GHz to 17 GHz are 0.8878, 0.9048, 0.9088, 0.9099, 0.9085, and 0.8959, respectively. It is observed that the quality of the reconstructed images performs best at the target frequency points while the nearby frequency crosstalk exhibits worse quality.

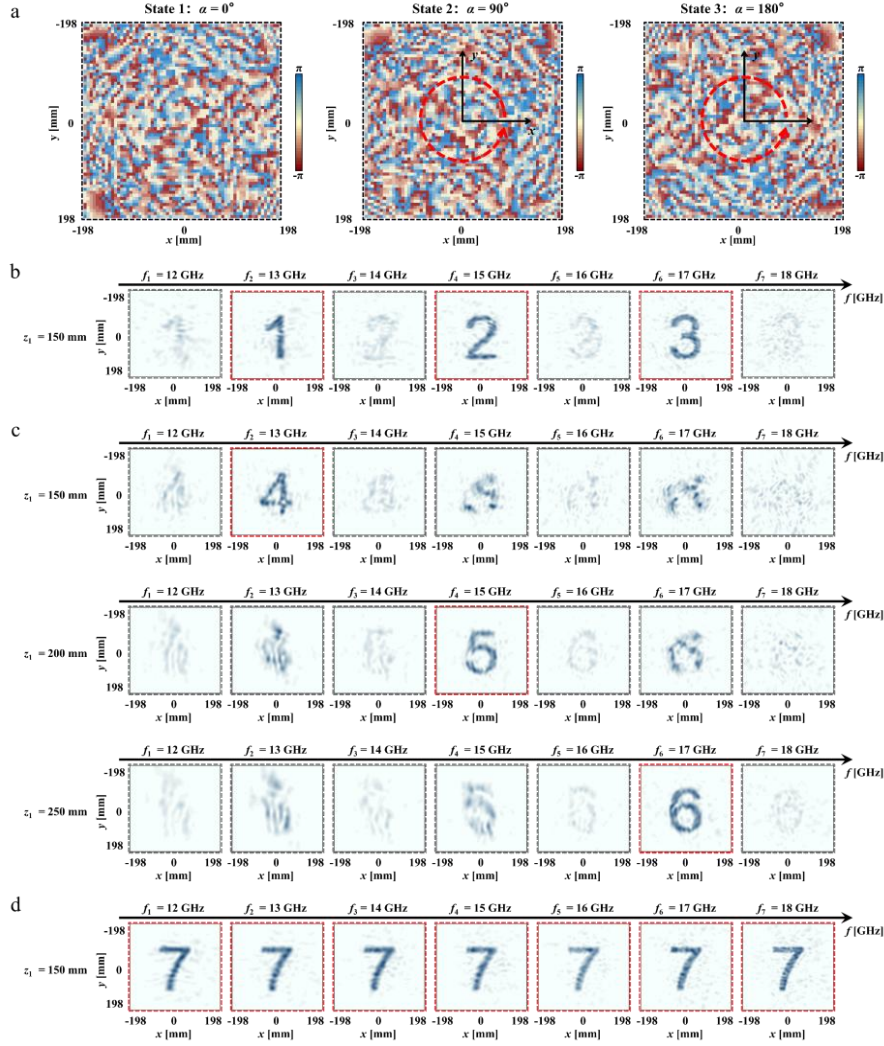

**Fig. 4** The simulated performance of the proposed layered twisted metasurface system. a) The matched phase distribution of PM at state 1, state 2 and state 3 with twisted angle of $\alpha = 0°$, $90°$ and $180°$. b) The simulated intensity distribution of state 1 for frequency ranging from 12 GHz to 18 GHz and focal $z$-axis plane $z_1 = 150$mm. c) The simulated intensity distribution of state 2 for frequency ranging from 12 GHz to 18 GHz and focal $z$-axis plane $z_1 = 150$mm, $z_2 = 200$mm, and $z_3 = 250$mm. d) The simulated intensity distribution of state 3 for frequency ranging from 12 GHz to 18 GHz and focal $z$-axis plane $z_1 = 150$mm.

Furthermore, a proof-of-concept meta-device prototype is fabricated by the printed circuit board etching technology. The prototype occupies the size of 396 mm × 396 mm. The RA-M is composed of two pieces of dielectric substrates with thicknesses of 1.93 mm and 0.43 mm respectively and one bonding layer with a thickness of 0.1 mm, as depicted in Fig. 5a(i-ii). The P-M is composed of two pieces of dielectric substrates with a thickness of both 1.5 mm and one bonding layer with a thickness of 0.1 mm, as depicted in Fig. 5a(iii-iv). A coaxial terminal is soldered to the end of the RA-M feeding network to provide the integrated excitation. To ensure the distance between the twisted metasurfaces as well as their alignment, we use a 3D-printed resin holder to fix the whole prototype. The measurement is performed in the microwave anechoic chamber. One section of the vector network analyzer is connected to the feed coaxial terminal of the RA-M, and the other end is connected to the near-field probe antenna. The near-field probe antenna is fixed to the scanning turntable to enable near-field detection of the plane to be measured. By measuring the near-field amplitude and phase of a pair of orthogonal polarizations, a CP target image can eventually be synthesized. The measured results are presented in Fig. 5b-d. It is shown that as the rotation of P-M, the switch function of the proposed meta-device is achieved and the measured intensity distribution agrees well with the target setting. For each state, the intensity distribution at a specific frequency and focal $z$-axis plane performs dispersion-customized properties. The measured imaging efficiency for state 1 of "1", "2", and "3" are 13.21%, 20.86%, and 32.68%, respectively. The measured imaging efficiency for state 2 of "4", "5", and "6" are 20.01%, 22.48%, and 23.47%, respectively. The measured imaging efficiency for state 3 of "7" ranging from 12 GHz to 18 GHz is 11.68%, 13.13%, 11.46%, 12.31%, 14.32%, and 13.65%, respectively. We also calculated the SNR and PCC to evaluate the image quality. The measured SNR for state 1 of "1", "2", and "3" are 1.84 dB, 2.91 dB, and 3.08 dB, respectively. The measured

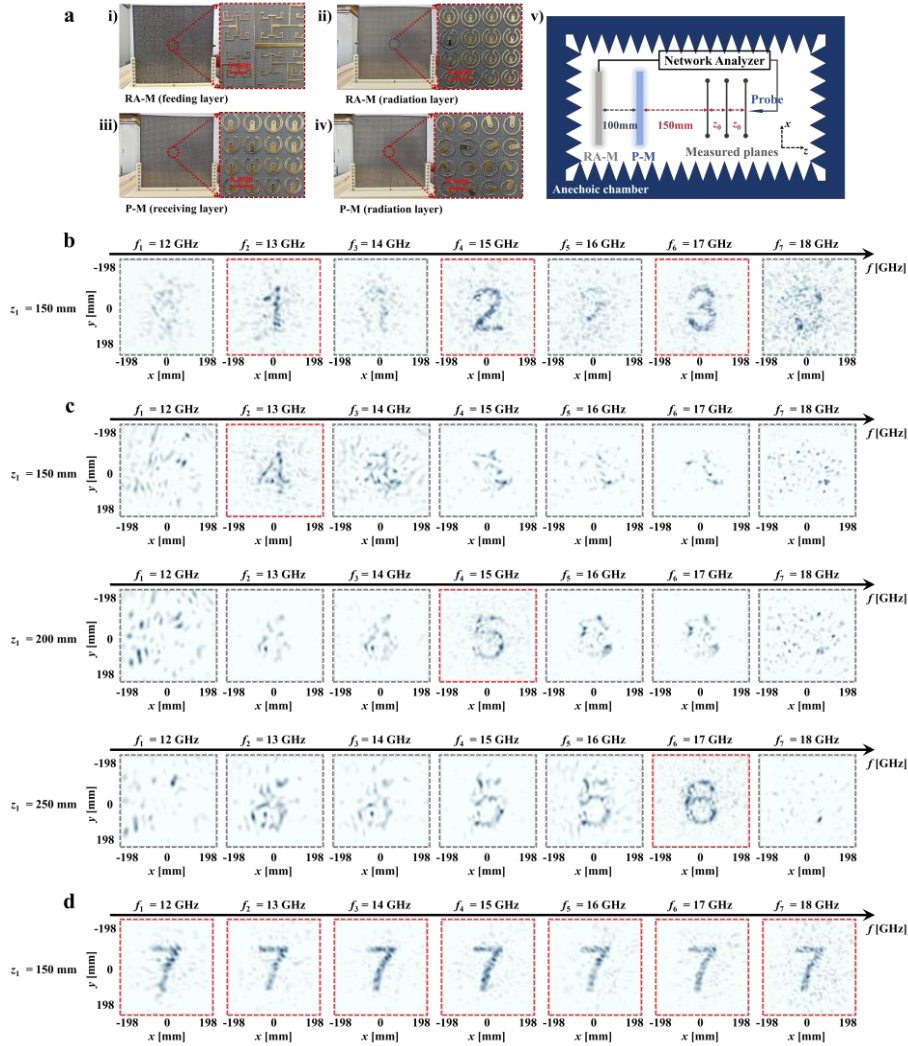

**Fig. 5** Experimental fabrication and measurement of the proposed layered twisted metasurface system. a) The photographs of the proposed meta-device prototype covering i) the bottom view of the RA-M; ii) the top view of the RA-M; iii) the bottom view of the P-M; iv) top view of the P-M, and v) the experimental setup in the microwave anechoic chamber for near-field intensity measurement. b) The measured intensity distribution of state 1 for frequency ranging from 12 GHz to 18 GHz and focal z-axis plane $z_1$ = 150mm. b) The measured intensity distribution of state 2 for frequency ranging from 12 GHz to 18 GHz and focal z-axis plane $z_1$ = 150mm, $z_2$ = 200mm, and $z_3$ = 250mm. c) The measured intensity distribution of state 3 for frequency ranging from 12 GHz to 18 GHz and focal z-axis plane $z_1$ = 150mm.

SNR for state 2 of "4", "5", and "6" are 1.96 dB, 2.88 dB, and 2.93 dB, respectively. The measured SNR for state 3 of "7" ranging from 12 GHz to 18 GHz is 2.85 dB, 3.01 dB, 3.13 dB, 2.82 dB, 2.54 dB, and 2.03 dB, respectively. The measured PCC for state 1 of "1", "2", and "3" are 0.8130, 0.8485, and 0.7982, respectively. The measured PCC for state 2 of "4", "5", and "6" are 0.5130, 0.5471, and 0.5962, respectively. The measured PCC for state 3 of "7" ranging from 12 GHz to 18 GHz are 0.8452, 0.8945, 0.8796, 0.8932, 0.8657, and 0.8839, respectively. Generally, the proposed dispersion-engineered inverse design framework can achieve the dispersion reconstruction engineering design of the frequency-reconfigurable holograms.

**Conclusions**

To conclude, we propose a dispersion-engineered inverse design framework based on the layered configuration of cascaded metasurfaces to realize the frequency-reconfigurable holography. The compact twisted metasurface system is composed of an integrated feeding RA-M and a rotational P-M. The RA-M is used to provide the modulated excitation for P-M while the in-plane rotation of P-M achieves the dynamic switching of the holograms. The inverse design framework is applied to optimize the phase profile of both RA-M and P-M to realize high-quality reconstruction of the target dispersion-customized holographic images. A proof-of-concept meta-device prototype is fabricated and measured to confirm the validity of our method. The switching of 3D frequency-space multiplexing holography and achromatic holography are demonstrated in the microwave region. Our framework provides new insights for dispersion engineering in metasurface-based holography and can be expanded to the entire spectrum and other dispersion-related applications.

**Abbreviations**

RA-M    Radiation-type metasurface

P-M     Phase-only metasurface

3D      three-dimensional

| | |
|---|---|
| EM | electromagnetic |
| DIDF | dispersion-engineered inverse design framework |
| MSE | mean squared error |
| CP | circularly polarized |
| SNR | signal-to-noise ratio |
| PCC | Pearson Correlation Coefficient |

## Supporting Information

Supporting Information is available.

## Declarations

**Acknowledgments**

Not applicable.

**Authors' contributions**

C. P. and Y. W. contributed equally to this work. J. Q. and Y. D. proposed the original idea and supervised the project. C. P. and Y. W. conducted the theoretical and simulation analyses. P. W., A. Y., and Y. L. fabricated the samples and performed the measurements. C. P., X. Y., M. H., and J. H. wrote the manuscript. All authors have approved the final version of the manuscript.

**Funding**

We are grateful for financial support from the National Natural Science Foundation of China (Grant Nos. 61901242, 62271170) and the National Key Laboratory of Laser Spatial Information Foundation.

**Availability of data and materials**

The datasets used and analyzed during the current study are available from the corresponding author upon reasonable request.

**Competing interests**

The authors declare that they have no competing interests

# References

1.  Li X, Chen QM, Zhang X, Zhao RZ, Xiao SM, et al. Time-sequential color code division multiplexing holographic display with metasurface. Opto-Electron Adv. 2023;6:220060;
2.  Zhang Z, Wu L. Domain multiplexed computer-generated holography by embedded wavevector filtering algorithm. PhotoniX. 2021;2:1.
3.  Xu C, Zhao RZ, Zhang X, et al. Quasicrystal metasurface for dual functionality of holography and diffraction generation. eLight. 2024;4:9.
4.  Wang YZ, Yu AX, Cheng YY, Qi JR. Matrix Diffractive Deep Neural Networks Merging Polarization into Meta-Devices. Laser Photonics Rev. 2024;18:2300903.
5.  Guo XY, et al. Stokes meta-hologram toward optical cryptography. Nat. Commun. 2022;13:6687.
6.  Wang RX, et al. Compact multi-foci metalens spectrometer. Light Sci. Appl. 2023;12:103.
7.  Chen C, et al. Spectral tomographic imaging with aplanatic metalens. Light Sci. Appl. 2023;8:99.
8.  Cai GY, et al. Compact angle-resolved metasurface spectrometer. Nat. Mater. 2024;23:71-78.
9.  Li JX, Li XR, Yardimci NT, Hu JT, Li YH, Chen JJ, Hung YC, Jarrahi M, Ozcan A. Rapid sensing of hidden objects and defects using a single-pixel diffractive terahertz sensor. Nat. Commun. 2023;14:6791.
10. Cotrufo M, Arora A, Singh S, Alù A. Dispersion engineered metasurfaces for broadband, high-NA, high-efficiency, dual-polarization analog image processing. Nat. Commun. 2023;14:7078.
11. Choi M, et al. Roll-to-plate printable RGB achromatic metalens for wide-field-of-view holographic near-eye displays. Nat. Mater. 2025.
12. Lee S, Jo Y, Yoo D, Cho J, Lee D, Lee B. Tomographic near-eye displays. Nat. Commun. 2019;10:2497.
13. Intaravanne Y, Wang RX, Ahmed H, Ming Y, Zheng YQ, Zhou ZK, Li ZC, Chen SQ, Zhang S, Chen XZ. Color-selective three-dimensional polarization structures. Light Sci. Appl. 2022;11:303.
14. So S, Kim J, Badloe T, Lee C, Yang YH, Kang H, Rho J. Multicolor and 3D Holography Generated by Inverse-Designed Single-Cell Metasurfaces. Adv. Mater. 2023;35:2208520.
15. Li J, Li XX, Huang XY, Kaissner R, Neubrech F, Sun S, Liu N. High Space-Bandwidth-Product (SBP) Hologram Carriers Toward Photorealistic 3D Holography. Laser Photonics Rev. 2024;28:2301173.
16. Wang YZ, Wang YF, Yu AX, Hu MS, Wang QM, Pang C, Xiong HM, Cheng YY, Qi JR. Non-Interleaved Shared-Aperture Full-Stokes Metalens via Prior-Knowledge-Driven Inverse Design. Adv. Mater. 2025;27:2408978.
17. Zhang Q, et al. Broadband Rapid Polarization Manipulation and Imaging Based on Pancharatnam-Berry Optical Elements. Laser Photonics Rev. 2025;19:2401536.
18. Mu XY, Qin HY, Zhao WN, Han SY, Liu ZQ, Shi YZ, Huang W, Li B, Song QH. Chirality-Free Full Decoupling of Jones Matrix Phase-Channels with a Planar Minimalist Metasurface. Nano Lett. 2025;25:1322-1328.
19. Yang H, Ou K, Liu Q, Peng MY, Xie ZW, Jiang YT, Jia HH, Cheng XB, Jing H, Hu YQ, Duan HG. Metasurface higher-order poincaré sphere polarization detection clock. Light Sci. Appl. 2025;14:63.
20. Hu JT, Mengu D, Tzarouchis DC, Edwards B, Engheta N, Ozcan A. Diffractive optical computing in free space. Nat. Commun. 2024;15:1525.
21. Chen WT, Zhu AY, Capasso F. Flat optics with dispersion-engineered metasurfaces. Nat. Rev. Mater. 2020;5:604-620.
22. Arbabi E, Arbabi A, Kamali SM, Horie Y, Faraon A. Multiwavelength polarization-

insensitive lenses based on dielectric metasurfaces with meta-molecules. Optica 2016;3:628.
23. Li K, Guo YH, Pu MB, Li X, Ma XL, Zhao ZY, Luo XG. Dispersion controlling meta-lens at visible frequency. Opt. Express. 2017;25:21419-21427.
24. Li JX, Yuan YY, Yang GH, Wu Q, Zhang W, Burokur SN, Zhang K. Hybrid Dispersion Engineering based on Chiral Metamirror. Laser Photonics Rev. 2023;17:2200777.
25. Aieta F, Kats MA, Genevet P, Capasso F. Multiwavelength achromatic metasurfaces by dispersive phase compensation. Science. 2015;347:1342-134.
26. Wang SM, et al. Broadband achromatic optical metasurface devices. Nat. Commun. 2017;8:187.
27. Ni XJ, Kildishev AV, Shalaev VM. Metasurface holograms for visible light. Nat. Commun. 2013;4:2807.
28. Chen QK, Gao YB, Pian SJ, Ma YG. Theory and Fundamental Limit of Quasiachromatic Metalens by Phase Delay Extension. Phys. Rev. Lett. 2023;131:19380.
29. Chen WT, Zhu AY, Sanjeev V, Khorasaninejad M, Shi ZJ, Lee Eric, Capasso F. A broadband achromatic metalens for focusing and imaging in the visible. Nat. Nanotechnol. 2018;13:220-226.
30. Wang YJ et al. High-efficiency broadband achromatic metalens for near-IR biological imaging window. Nat. Commun. 2021;12:5560.
31. Shrestha S, Overvig AC, Lu M, Stein A, Yu NF. Broadband achromatic dielectric metalenses. Light Sci. Appl. 2018;7:85.
32. Avayu O, Almeida E, Prior Y, Ellenbogen T. Composite functional metasurfaces for multispectral achromatic optics. Nat. Commun. 2017;8:14992.
33. Ye WM et al. Spin and wavelength multiplexed nonlinear metasurface holography. Nat. Commun. 2016;7:11930.
34. Jang J, Moon S-W, Kim J, Mun J, Maier SA, Ren H, Rho J. Wavelength-multiplexed orbital angular momentum meta-holography. PhotoniX. 2024;5:27.
35. Wang JX, Yu FL, Chen J, Wang J, Chen RS, Zhao ZY, Chen J, Chen XS, Lu W, Li GH. Continuous-spectrum–polarization recombinant optical encryption with a dielectric metasurface. Adv. Mater. 2023;35:2304161.
36. Chi HB, Hu YQ, Ou XN, Jiang YT, Yu D, Lou SZ, Wang Q, Xie Q, Qiu C-W, Duan HG. Neural Network-Assisted End-to-End Design for Full Light Field Control of Meta-Optics. Adv. Mater. 2025;2419621.
37. Deng ZL et al. Full-Color Complex-Amplitude Vectorial Holograms Based on Multi-Freedom Metasurfaces. Adv. Funct. Mater. 2020;30:1910610.
38. Li X, Chen LW, Li Y, Zhang XH, Pu MB, Zhao ZY, Ma XL, Wang YQ, Hong MH, Luo XG. Multicolor 3D meta-holography by broadband plasmonic modulation. Sci. Adv. 2017;2:e1601102.
39. Hu YQ, Jiang YT, Zhang Y, Yang X, Ou XN, Li L, Kong XH, Liu XS, Qiu C-W, Duan HG. Asymptotic dispersion engineering for ultra-broadband meta-optics. Nat. Commun. 2023;14:6649.
40. Li JX, Kamin S, Zheng GX, Neubrech F, Zhang S, Liu N. Addressable metasurfaces for dynamic holography and optical information encryption. Sci. Adv. 2018;4:eaar6768.
41. Shi L, Li BC, Kim C, Kellnhofer P, Matusik W. Towards real-time photorealistic 3D holography with deep neural networks. Nature. 2021;591:234-239.
42. Ou GY, Yang WH, Song QH, Liu YL, Qiu C-W, Han JC, Tsai D-P, Xiao SM. Reprogrammable meta-hologram for optical encryption. Nat. Commun. 2020;11:5484.
43. Fan ZX, Qian C, Jia YT, Feng YM, Qian HL, Li E-P, Fleury R, Chen HS. Holographic multiplexing metasurface with twisted diffractive neural network. Nat. Commun. 2024;15:9416.
44. Li JX, Yu P, Zhang S, Liu N. Electrically-controlled digital metasurface device for light projection displays. Nat. Commun. 2020;11:3574.
45. Li WZ, Yu QY, Qiu JH, Qi JR. Intelligent wireless power transfer via a 2-bit compact

reconfigurable transmissive-metasurface-based router. Nat. Commun. 2024;15:2807.
46. Cai, X. et al. Dynamically controlling terahertz wavefronts with cascaded metasurfaces. Adv. Photon. 2021;3:036003.
47. Wang YH, Pang C, Wang YZ, Qi JR. Electromagnetic Manipulation Evolution from Stacked Meta-Atoms to Spatially Cascaded Metasurfaces. Ann. Phys. (Berlin) 2025;537:2400158.
48. Li HM, Yu AX, Pang C, Wang YZ, Qi JR. Full-Phase Parameter Modulation with Arbitrary PolarizationCombination via Bidirectional Asymmetric TransmissionMeta-Devices. Laser Photonics Rev. 2024;18:2400300.
49. Wang YZ, Pang C, Qi JR. 3D Reconfigurable Vectorial Holography via a Dual-Layer Hybrid Metasurface Device. Laser Photonics Rev. 2024;18:230083.
50. Zhang M et al. Plasmonic Metasurfaces for Switchable Photonic Spin-Orbit Interactions Based on Phase Change Materials. Adv. Sci. 2018;5:1800835.
51. Wang ZP, Jiang L, Li XW, Li BH, Zhou SP, Xu ZT, Huang LL. Thermally Reconfigurable Hologram Fabricated by Spatially Modulated Femtosecond Pulses on a Heat-Shrinkable Shape Memory Polymer for Holographic Multiplexing. ACS Appl. Mater. Interfaces 2021;13:51736−51745.
52. Mu YJ, Xia DX, Han JQ, Ma XG, Wang X, Liu HX, Li L. Time-Space-Coding Radiation-Stealth Metasurface with Amplitude-Phase Co-Modulation. Adv. Funct. Mater. 2024;34:2407802.
53. Liu W, Wang SR, Dai JY, Zhang L, Cheng Q, Cui TJ. Arbitrarily rotating polarization direction and manipulating phases in linear and nonlinear ways using programmable metasurface. Light Sci. Appl. 2024;13:172.
54. Hu Q et al. Joint Amplitude-Phase Metasurface for Polarization-Selective Dynamic Wavefront Manipulation and Broadband Absorption. Adv. Mater. Technol. 2023;8:2300111.